\documentclass[superscriptaddress,twocolumn,showpacs,preprintnumbers,amsmath,amssymb,prc]{revtex4}

\usepackage{graphicx}
\usepackage{subfigure}
\usepackage{dcolumn}
\usepackage{bm}

\begin{document}


\title{Nuclear structure study of $^{19,20,21}$N nuclei by $\gamma$ spectroscopy}

\author{Z.~Elekes}
\affiliation{Institute of Nuclear Research of the
Hungarian Academy of Sciences, P.O. Box 51, Debrecen, H-4001, Hungary}
\author{Zs.~Vajta}
\affiliation{Institute of Nuclear Research of the
Hungarian Academy of Sciences, P.O. Box 51, Debrecen, H-4001, Hungary}
\author{Zs.~Dombr\'adi}
\affiliation{Institute of Nuclear Research of the
Hungarian Academy of Sciences, P.O. Box 51, Debrecen, H-4001, Hungary}
\author{T.~Aiba}
\affiliation{Niigata University, Niigata 950-2181, Japan}
\author{N.~Aoi}
\affiliation{The Institute of Physical and
Chemical Research (RIKEN), 2-1 Hirosawa, Wako, Saitama 351-0198, Japan}
\author{H.~Baba}
\affiliation{The Institute of Physical and
Chemical Research (RIKEN), 2-1 Hirosawa, Wako, Saitama 351-0198, Japan}
\author{D.~Bemmerer}
\affiliation{Forschungszentrum Dresden-Rossendorf, Bautzner Landstrasse 400, 01328 Dresden (Rossendorf), Germany}
\author{Zs.~F\"ul\"op}
\affiliation{Institute of Nuclear Research of the
Hungarian Academy of Sciences, P.O. Box 51, Debrecen, H-4001, Hungary}
\author{N.~Iwasa}
\affiliation{Tohoku University, Sendai, Miyagi 9808578, Japan}
\author{\'A.~Kiss}
\affiliation{E\"otv\"os Lor\'and University, 1117 Budapest,
P\'azm\'any P\'eter s\'et\'any 1/A, Hungary}
\author{T.~Kobayashi}
\affiliation{Tohoku University, Sendai, Miyagi 9808578, Japan}
\author{Y.~Kondo}
\affiliation{Tokyo Institute of Technology,
2-12-1 Oh-Okayama, Meguro, Tokyo 152-8551, Japan}
\author{T.~Motobayashi}
\affiliation{The Institute of Physical and
Chemical Research (RIKEN), 2-1 Hirosawa, Wako, Saitama 351-0198, Japan}
\author{T.~Nakabayashi}
\affiliation{Tokyo Institute of Technology,
2-12-1 Oh-Okayama, Meguro, Tokyo 152-8551, Japan}
\author{T.~Nannichi}
\affiliation{Tokyo Institute of Technology,
2-12-1 Oh-Okayama, Meguro, Tokyo 152-8551, Japan}
\author{H.~Sakurai}
\affiliation{The Institute of Physical and
Chemical Research (RIKEN), 2-1 Hirosawa, Wako, Saitama 351-0198, Japan}
\author{D.~Sohler}
\affiliation{Institute of Nuclear Research of the
Hungarian Academy of Sciences, P.O. Box 51, Debrecen, H-4001, Hungary}
\author{S.~Takeuchi}
\affiliation{The Institute of Physical and
Chemical Research (RIKEN), 2-1 Hirosawa, Wako, Saitama 351-0198, Japan}
\author{K.~Tanaka}
\affiliation{The Institute of Physical and
Chemical Research (RIKEN), 2-1 Hirosawa, Wako, Saitama 351-0198, Japan}
\author{Y.~Togano}
\affiliation{The Institute of Physical and
Chemical Research (RIKEN), 2-1 Hirosawa, Wako, Saitama 351-0198, Japan}
\author{K.~Yamada}
\affiliation{The Institute of Physical and
Chemical Research (RIKEN), 2-1 Hirosawa, Wako, Saitama 351-0198, Japan}
\author{M.~Yamaguchi}
\affiliation{The Institute of Physical and
Chemical Research (RIKEN), 2-1 Hirosawa, Wako, Saitama 351-0198, Japan}
\author{K.~Yoneda}
\affiliation{The Institute of Physical and
Chemical Research (RIKEN), 2-1 Hirosawa, Wako, Saitama 351-0198, Japan}

\date{\today}

\begin{abstract}
The structure of neutron rich nitrogen nuclei has been studied by use of neutron removal reaction and inelastic scattering.
Mass and charge deformations have been deduced for the first excited state of $^{21}$N, which indicates the partial persitence of
the $N$=14 subshell closure in nitrogen isotopes. The spectroscopic information obtained on the structure of $^{19,20,21}$N
confirms the results from a previous experiment.
\end{abstract}

\pacs{23.20.Js, 25.60.-t, 27.30.+t, 29.30.Kv}
\maketitle

\section{Introduction}
Due to the proton-neutron monopole interaction, the energies of the single particle states are changing along an isotopic or
isotonic chain by varying the number of neutrons or protons, respectively. As a consequence, shell gaps may disappear,
or new shell closures may develop. One of the new subshell closures is at $N$=14~\cite{thirolf,becheva,22o}
the strength of which was determined to be 4.2 MeV in $^{22}$O~\cite{stanoiu_22o}. Recently, disappearence of this $N$=14 gap
in carbon isotopes has been reported via the spectroscopy of $^{20}$C~\cite{stanoiu_prc}. This observation has been confirmed
also by the measurement of the mass deformation in $^{20}$C~\cite{elekes_20c} in inelastic scattering on a liquid hydrogen target.
As a byproduct of this study, we investigated the interaction of the $^{21}$N beam with the hydrogen and a lead target, as well.
This allowed us to study the transiton along the $N$=14 line and to give spectroscopic information on the structure of
the lighter neutron rich nitrogen isotopes.

\section{Experimental details}
\begin{figure}[]
\includegraphics[scale=0.4,clip=]{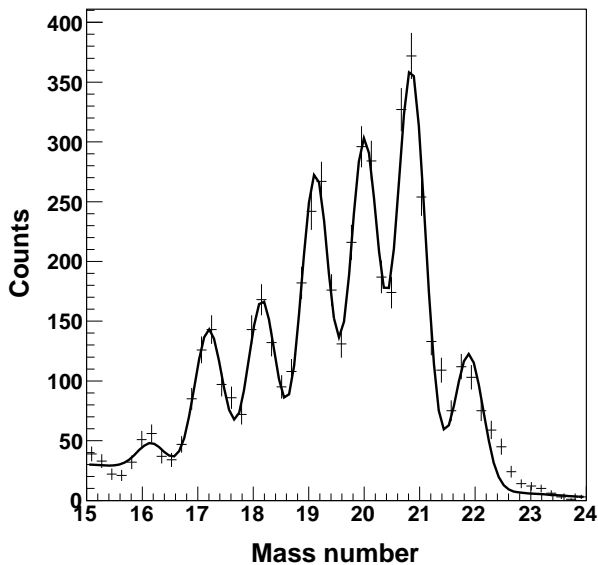}
\caption{Linearized mass distribution of nitrogen isotopes from the $^{21,22}$N+$^1$H collision.}
\label{fig:aid}
\end{figure}

The experiment was performed at RIKEN Nishina Center using a $^{40}$Ar primary beam of
63~MeV/nucleon energy and 700~pnA intensity. The radioactive species were produced
through fragmentation reaction on a $^{181}$Ta target of 0.2~mm thickness,
and their momentum and mass were analyzed by the RIKEN isotope separator (RIPS)~\cite{rips}.
An aluminium wedge degrader of 221~mg/cm$^2$ thickness was placed at the momentum dispersive focal plane (F1) in
order to purify the secondary beam. RIPS momentum acceptance was set to the maximum value of 6\%
in order to achieve the highest intensities since a sharply defined energy
was not crucial for our purposes. The resultant beam included several nuclei, mainly
$^{17}$B, $^{19}$C, $^{20}$C, $^{21}$N and $^{22}$N with a total intensity of about 100~particle/s (pps),
to which the contribution of $^{21}$N was around 30~pps.
The identification of these constituents was carried out event-by-event using
energy loss ($\Delta$E), time-of-flight (ToF) and magnetic rigidity ($B\rho$) information~\cite{sakurai}.
A complete separation of the istopes was achieved by the $\Delta$E-ToF-$B\rho$ method.
Two plastic scintillators of 0.3~mm thickness placed at the second and third
focal planes (F2 and F3) determined the ToF, while the $\Delta$E value was measured by a Si detector
of 0.1~mm thickness at F2. PPACs at F2 and F3 monitored the cocktail beam which was transported to the
secondary targets of $^{208}$Pb and liquid hydrogen with 1445~mg/cm$^2$ and 190~mg/cm$^2$ thickness, respectively.
The mean energy of the $^{21}$N ions in the middle of the hydrogen and lead target was
52.0~MeV/nucleon and 48.1~MeV/nucleon.
The scattered particles were detected and identified on the basis of $\Delta$E, ToF and total energy (E)
information. 80~cm downstream of the target a plastic scintillator of 1~mm was put to measure the $\Delta$E
and to provide the start signal for the ToF measurement. The end of the ToF signal was determined
by an array of plastic detectors located 4.3~m downstream of the target, which also measured the total
energy of the isotopes. This array consisted of 16 bars of 60~mm thickness with a total area of 1$\times$1~m$^2$,
and its angular acceptance of 6.5$^\circ$ in the laboratory frame granted almost 100\% coverage of the cross section.
160 NaI(Tl) crystals of the DALI2 array~\cite{dali} surrounding the targets detected the
$\gamma$ rays emitted by the inelastically scattered nuclei.
Since knock-out reaction channels were strong for the hydrogen target (for Pb target this effect was negligible),
the separation of the outgoing isotopes was necessary. The identification of the atomic number (Z)
was perfect using the $\Delta$E and ToF information, while the mass number (A) was determined
on the basis of ToF and total energy (Fig.~\ref{fig:aid}).
The segregation in A for the segments of the plastic
scintillator array was different therefore only those detectors were used for the identification
of $\gamma$ rays where the adjacent mass numbers were completely resolved. However, for the deduction
of cross section of $^{21}$N inelastic scattering reaction, the other detectors were also employed because
the $^{21}$N and $^{19}$N nuclei, which emitted $\gamma$ rays with similar energy, were distinct.

\section{Results}
\subsection{Level scheme of neutron-rich nitrogen isotopes}
The Doppler-corrected spectra using the hydrogen target were produced with multiplicity (M) 1 and 2 of the DALI2 array
for all the isotopes. Spectra for $^{18}$N served as a cross-check of the procedure, since
the low-energy excited states for this nucleus are well-known~\cite{18n}. Using this information
a \textsc{Geant4}~\cite{geant4} simulation of the M=1 and M=2 spectra was performed and
the resulting response curves plus additional smooth background polynomials fitted the
experimental data points well.
For the heavier isotopes, the spectra were fitted first by Gaussian functions
and polynomial backgrounds in order to determine the position of the peaks. These were then
compared to the $\gamma$ lines and level scheme found in Ref.~\cite{sohler}. The resulting
peaks were fed into \textsc{Geant4} simulation again and the response curves were fitted to the
experimental spectra and the relative intensities were deduced.

\begin{figure}[]
\subfigure{\includegraphics[scale=0.4,clip=]{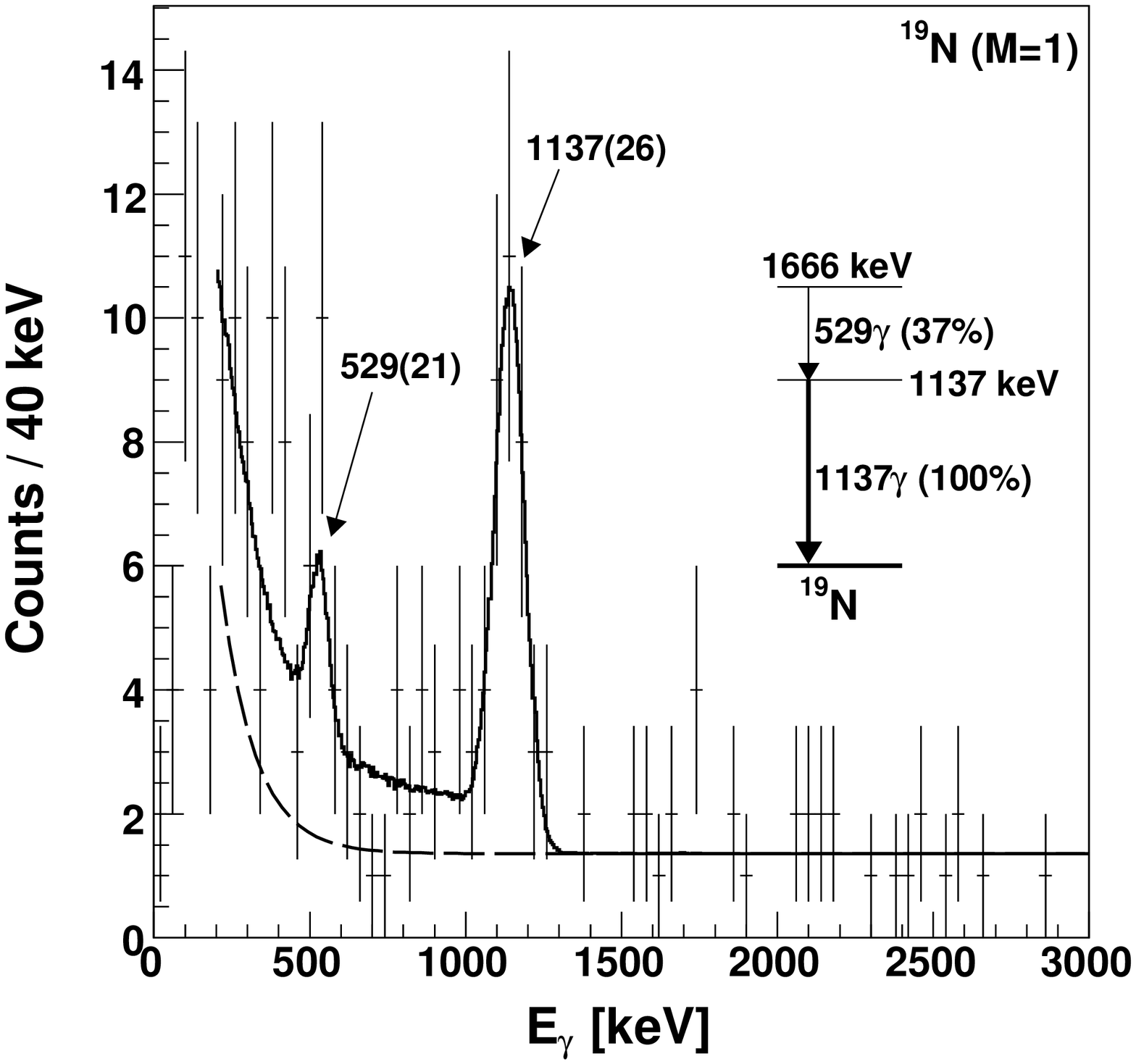}} \\
\subfigure{\includegraphics[scale=0.4,clip=]{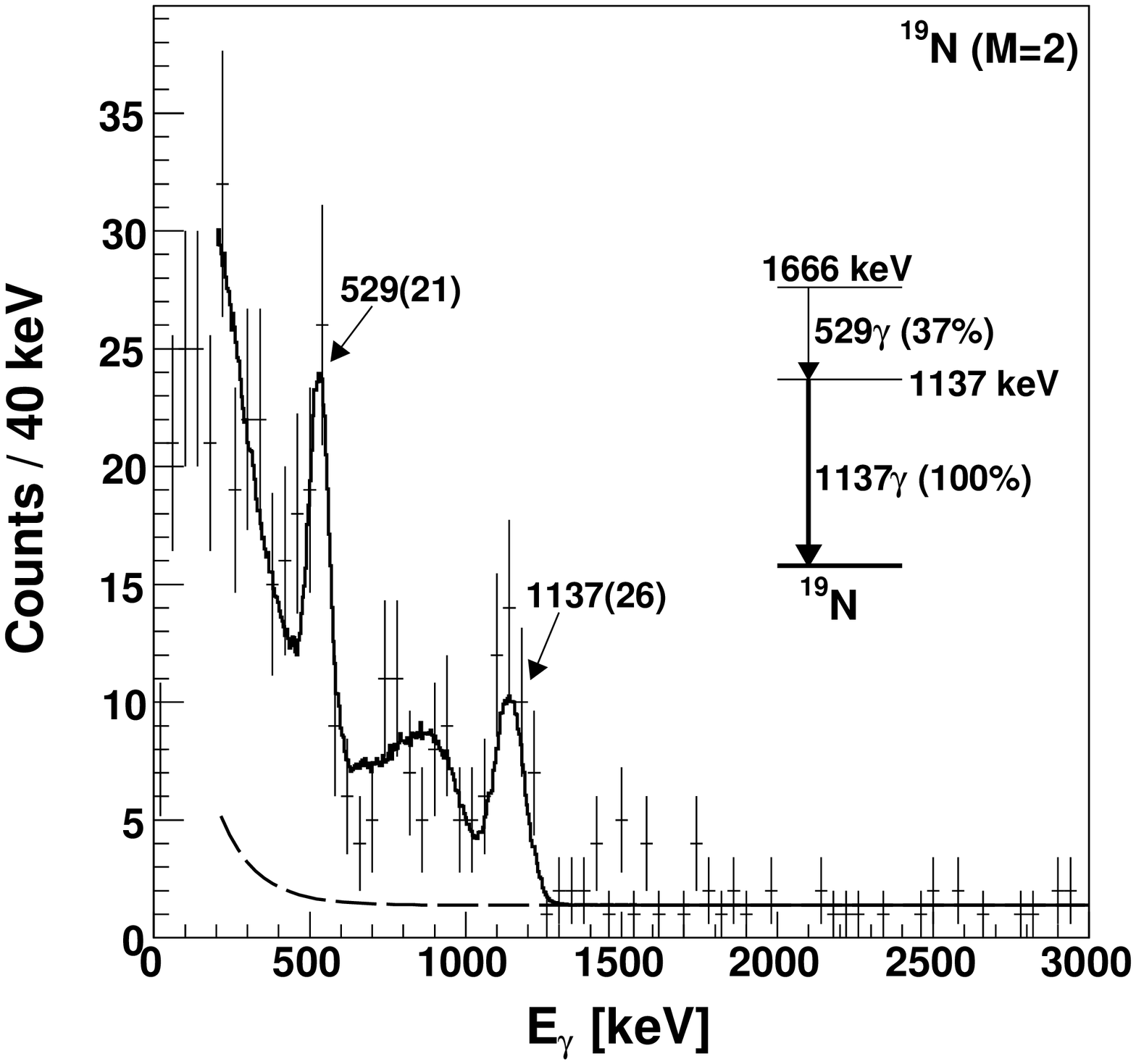}}
\caption{Doppler-corrected spectra of $^{19}$N $\gamma$ rays emerging from
neutron knock-out reactions with M=1 (upper panel) M=2 (lower panel).
The solid line is the final fit including the spectrum curves from \textsc{Geant4} simulation
and additional smooth polynomial backgrounds plotted
as separate dotted lines.}
\label{fig:19n}
\end{figure}

Two peaks were found for $^{19}$N in the neutron knock-out reaction at 529(21)~keV and 1137(26)~keV (Fig.~\ref{fig:19n}).
This is in a good agreement with Ref.~\cite{sohler} where these transition form a cascade.
The lower energy $\gamma$ ray connects the second excited state with the first one while
the higher energy $\gamma$ ray is emitted during the transition from the first excited state
to the ground state.

The $^{20}$N neutron knock-out spectra are more complicated and could be best fitted by four peaks at
500(21)~keV, 832(17)~keV, 920(30)~keV and 1052(24)~keV. This conclusion is mainly based on the
M=2 spectrum since the M=1 one contains basically the 832~keV peak.
This observation is consistent with the level scheme deduced in Ref.~\cite{sohler} which
consists of two cascades with similar energies.

\begin{figure}[]
\subfigure{\includegraphics[scale=0.4,clip=]{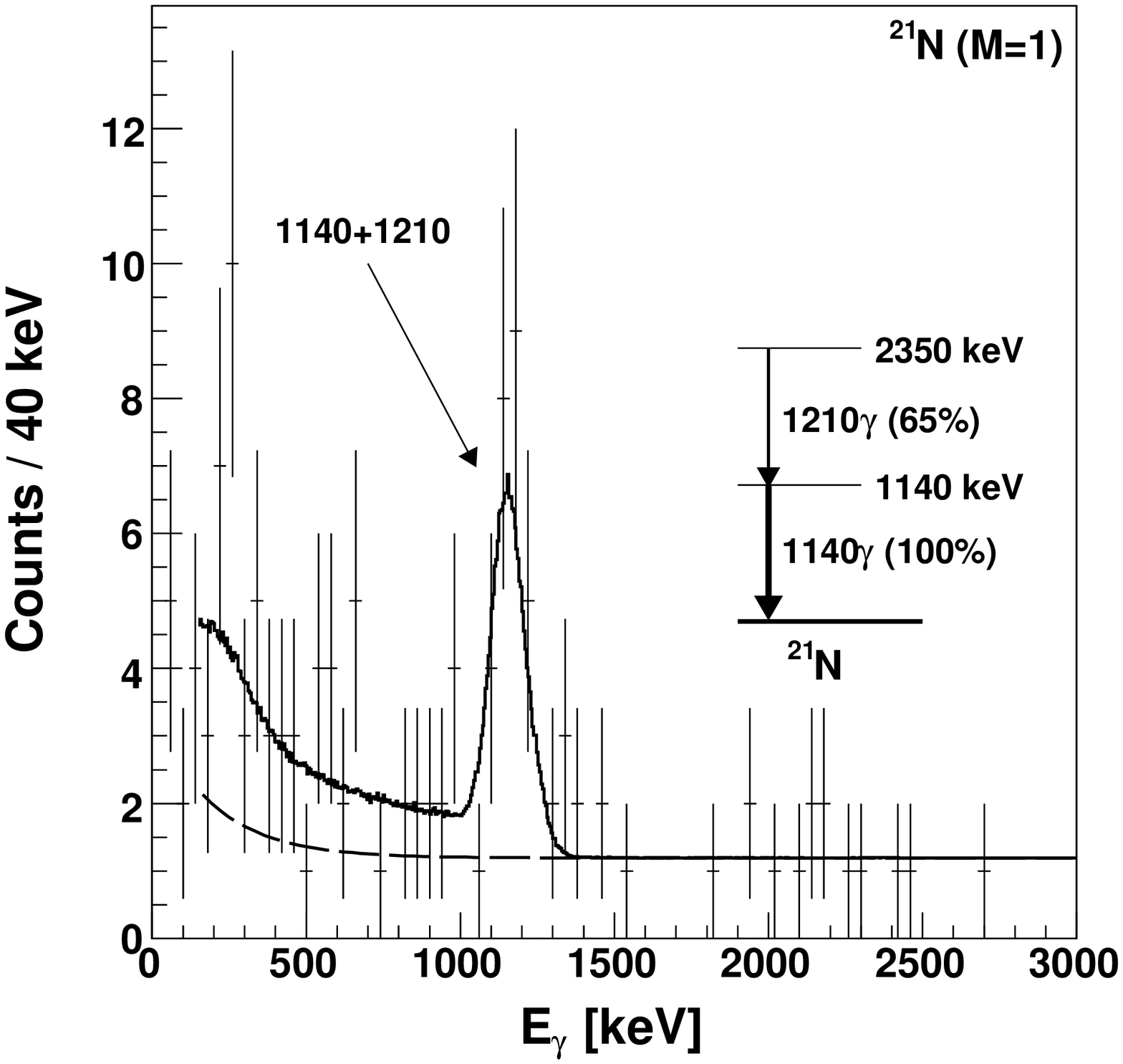}} \\
\subfigure{\includegraphics[scale=0.4,clip=]{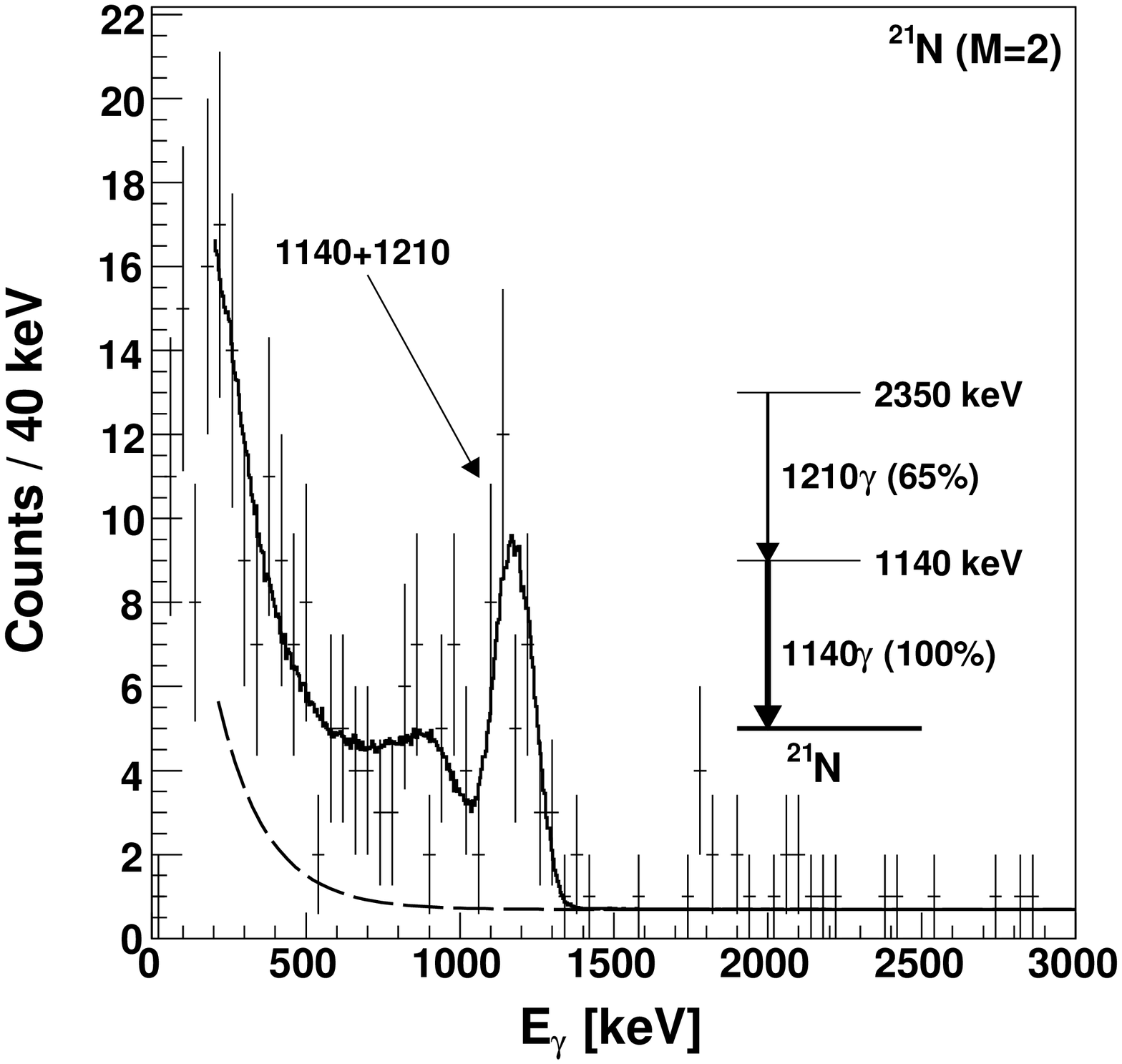}}
\caption{Doppler-corrected spectra of $\gamma$ rays emerging from
$^{21}$N+$^{1}$H reaction with M=1 (upper panel) and M=2 (lower panel).
The solid line is the final fit including the spectrum curves from \textsc{Geant4} simulation
and additional smooth polynomial backgrounds plotted
as separate dotted lines.}
\label{fig:21n}
\end{figure}

For $^{21}$N (Fig.~\ref{fig:21n}), the transition between the first excited and ground state (E$_\gamma$=1140(30)~keV) dominates
the proton inelastic scattering spectra with M=1 while the M=2 spectra also
contains events originating from the transition between the second and first excited state (E$_\gamma$=1210(33)~keV).
This agrees well with the conclusion drawn in Ref.~\cite{sohler} on the level scheme where the two strongest peaks
were found at 1177~keV and 1228~keV.

\subsection{Inelastic scattering cross sections for $^{21}$N}
Besides the level scheme, the inelastic scattering cross sections were also deduced
for $^{21}$N both with hydrogen and lead targets.
The net counts in the nearby peaks for the spectra with liquid hydrogen were derived by fixing the known peak positions
and the peak widths determined above. The relative intensities deduced from the spectrum containing all the multipolarities are:
100(10) for the 1140~keV and 60(6) for the 1210~keV line. The gamma feeding of the first excited state was taken into
account while deriving the excitation cross sections.
Since only the first excited state was populated
in the Pb case and no $\gamma$ ray was detected in the M=2 spectrum, no correction was taken into account there. 
The three excitation cross sections determined are $\sigma$(1140,H)=6.6(8)~mb, $\sigma$(2350,H)=12.3(16)~mb
and $\sigma$(1140,Pb)=11.4(34)~mb.

\section{Interpretation and Discussion}
As in several earlier cases, e.g.,~\cite{br_prc}, the cross sections were interpreted in terms of the collective
model using the \textsc{Fresco} code~\cite{fresco}.
The optical potential for the hydrogen target was calculated by a parametrized formula
from a global phenomenological evaluation~\cite{koning}, while the optical potential
determined experimentally in $^{17}$O+$^{208}$Pb reaction~\cite{pb_potential} was
applied for the lead target. The spins of the states involved
were taken from shell model calculations~\cite{sohler} as follows:
the ground state has a spin 1/2, the first excited state at 1140~keV has spin 3/2
while the second excited state at 2350~keV has spin 5/2.

\begin{figure}[]
\includegraphics[scale=0.4,clip=]{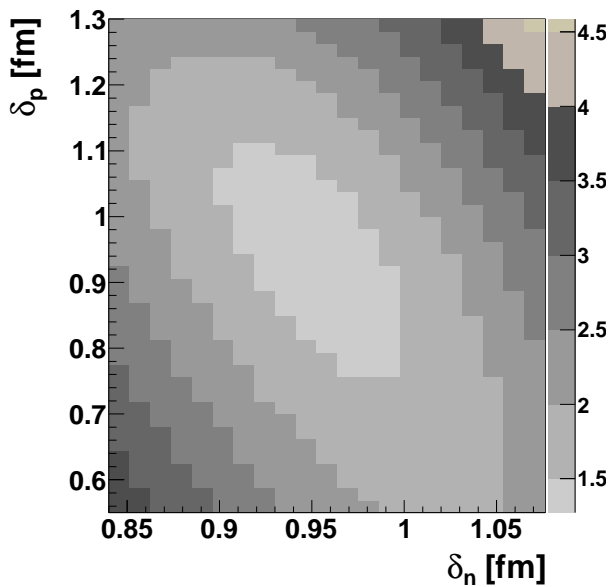}
\caption{
Contour plot of reduced $\chi^2$ values in function of the neutron and proton deformation lengths ($\delta_n$, $\delta_p$).}
\label{fig:chi}
\end{figure}

The determination of the proton and
neutron deformation lengths ($\delta_p$,$\delta_n$) was performed in a similar way
described in details in Ref.~\cite{br_prc} therefore we recall here some details.
The exception was that here the proton inelastic cross section
of the second excited state was also available and was used in the $\chi^2$ analysis.
As a first step, a pair of neutron and proton deformation lengths has been chosen. These are in the
following correspondence with the matter and Coulomb deformation lengths for the two probes
($\delta_M^{Pb}$, $\delta_M^{pp}$, $\delta_C^{Pb}$=$\delta_C^{pp}$=$\delta_p$):
\begin{equation}
(Z\cdot b_{p}^{Pb}+N\cdot b_{n}^{Pb})\cdot\delta_{M}^{Pb}
=N\cdot b_{n}^{Pb}\cdot\delta_{n}+Z\cdot b_{p}^{Pb}\cdot\delta_{p}
\end{equation}
\begin{equation}
(Z\cdot b_{p}^{pp}+N\cdot b_{n}^{pp})\cdot\delta_{M}^{pp}
=N\cdot b_{n}^{pp}\cdot\delta_{n}+Z\cdot b_{p}^{pp}\cdot\delta_{p}
\end{equation}
where $b_{n}^{Pb}$, $b_{p}^{Pb}$, $b_{n}^{pp}$ and $b_{p}^{pp}$
are the neutron and proton sensitivity parameters.
$\delta_{M,C}^{Pb}$, $\delta_{M,C}^{pp}$ are the input parameters in the coupled channel code.
The difference between the calculated and experimental cross sections has been quantified
in a $\chi^2$ value so we ended up with a set of data ($\delta_n$,$\delta_p$,$\chi^2$).
This procedure was repeated with varied initial ($\delta_n$,$\delta_p$) parameters
and the results are visualized in a contour plot of $\chi^2$ values (Fig.~\ref{fig:chi}).
From this figure, the neutron and proton deformation lengths can easily be determined
at $\delta_n$=0.95(5)~fm, $\delta_p$(=$\delta_C^{Pb}$=$\delta_C^{pp}$)=0.95(15)~fm
which implies $\delta_M^{Pb}$=$\delta_M^{pp}$=0.95(5)~fm.

The corresponding reduced
electric quadrupole transition probability ($B(E2)$) and the multipole proton and neutron transition matrix elements ($M_p,M_n$) could directly be
calculated with the following formulae:
\begin{equation}
B(E2;0_{gs}^{+}\rightarrow 2_{1}^{+})/e^2=M_{p}^{2}=(\frac{3}{4\pi}\cdot Z\cdot \delta_{p}\cdot\ R)^{2}
\label{mp2}
\end{equation}
\vspace{-0.7cm}
\begin{equation}
M_{n}^{2}=(\frac{3}{4\pi}\cdot N\cdot \delta_{n}\cdot\ R)^{2}
\label{mn2}
\end{equation}
at  M$_n^2$=110 (12)~fm$^4$, M$_p^2$=28 (9)~fm$^4$, and the ratio
$\frac{M_{n}}{M_{p}}$=2.0 (3). This ratio is equal to $N/Z$, which shows the pure isoscalar
character of the transition.  These results are similar to those
of the $^{22}$O nucleus having 14 neutrons just one proton closer to stability~\cite{becheva,22o}
where also small transition matrix elements together with $\frac{M_{n}}{M_{p}}=1.4(5)\frac{N}{Z}$
were extracted. 

According to the weak coupling approximation~\cite{alder}, the sum of the E2 strengths
from the ground state to the 3/2$^-$ and 5/2$^-$ states in the $^{21}$N isotope gives
the B(E2;$0^+\rightarrow 2^+$) strength in its appropriate  core.
We have measured the B(E2) value to the first excited state and the mass transition rate
to both the first and the second excited states. Since these states are expected to have mainly
$\pi p_{1/2}\oplus 2^+$ nature, it is a good approximation to assume that for both states
the ratio of the mass and electric transition rates is the same. Using this assumption,
the effective B(E2) value of the ground state transiton in the core of $^{21}$N can be estimated
as 56(18)~e$^2$fm$^4$. The shell model calculations give 54~e$^2$fm$^4$ summed strength
for $^{21}$N~\cite{sohler}, which is about twice of the $^{22}$O value and lies at about
half way between the neighbouring oxygen and carbon B(E2;$0^+\rightarrow 2^+$) values.
The shell model calculation gives 110~e$^2$fm$^4$ for $^{20}$C, which is much larger than
the experimental value ($<$18~e$^2$fm$^4$) due to the decoupling of the neutrons from the core
in heavy carbon nuclei~\cite{elekes_20c}. This fact shows that the core structure of the
nitrogen isotopes is softer than that of the singly-closed shell oxygen isotopes and
is consistent with the 1.2~MeV reduction of the $N$=14 shell closure when going from
$^{22}$O to $^{21}$N as a result of the removal of a proton from the $p_{1/2}$ orbit~\cite{sohler}.

\section{Acknowledgments}
We would like to thank the RIKEN Ring Cyclotron staff for their assist during
the experiment. One of authors (Z.~E.) is grateful for the Bolyai grant in Hungary.
The European authors thank the kind hospitality and support from RIKEN.
The present work was partly supported by the Grant-in-Aid for Scientific
Research (No. 1520417)
by the Ministry of Education, Culture, Sports, Science and Technology and by
OTKA K68801, T049837, NKTH JP-16/2006.

\end{document}